\documentclass[11pt]{article}
\usepackage{Moriond,epsfig}
\usepackage[usenames,dvipsnames]{color}
\usepackage{subfigure}
\RequirePackage[colorlinks=true
,urlcolor=blue
,anchorcolor=blue
,citecolor=blue
,filecolor=blue
,linkcolor=blue
,menucolor=blue
,pagecolor=blue
,linktocpage=true
,pdfproducer=medialab
]{hyperref}
\usepackage{cite}
\usepackage{amsmath}
\usepackage{mathrsfs}
\usepackage{amssymb}
\usepackage{slashed}

\newcommand{\red}[1]{\color{red} #1 \color{black}}
\newcommand{\blue}[1]{\color{blue} #1 \color{black}}
\newcommand{\green}[1]{\color{OliveGreen} #1 \color{black}}

\newcommand{\UH}{\mathbf{U}}

\newcommand{\TL}{\mathbf{T}}
\newcommand{\VL}{\mathbf{V}}
\newcommand{\DL}{\mathbf{D}}

\newcommand{\tmnp}{\sigma^{\mu\nu}}

\newcommand{\mean}[1]{\langle#1\rangle}
\newcommand{\derp}{\partial}

\newcommand{\LL}{\mathscr{L}}

\def\cD{{\cal D}}
\def\cF{{\cal F}}
\def\cO{{\cal O}}
\def\cP{{\cal P}}
\def\cX{{\cal X}}

\def\cY{{\bf Y}}
\def\cy{{\bf y}}
\def\Tr{{\rm Tr}}
\def\vep{\varepsilon}

\def\be{\begin{equation}}
\def\ee{\end{equation}}
\def\beq{\begin{equation}}
\def\eeq{\end{equation}}
\def\bc{\begin{center}}
\def\ec{\end{center}}
\def\bea{\begin{eqnarray}}
\def\eea{\end{eqnarray}}

\def\nn{\nonumber}

\def\vep{\varepsilon}

\def\be{\begin{equation}}
\def\ee{\end{equation}}
\def\bea{\begin{eqnarray}}
\def\eea{\end{eqnarray}}

\begin{document}
\vspace*{4cm}
\title{A LIGHT DYNAMICAL SCALAR BOSON}

\author{R.~Alonso$^1$, M.B.~Gavela$^{1,2}$, L.~Merlo$^{1,2,*}$, S.~Rigolin$^3$, J.~Yepes$^1$}

\address{\vspace{0.5cm}
$^1$Instituto de F\'{\i}sica Te\'orica UAM/CSIC and Departamento de F\'isica Te\'orica,\\
Universidad Aut\'onoma de Madrid, Cantoblanco, 28049 Madrid, Spain\\
$^2$CERN, Department of Physics, Theory Division CH-1211 Geneva 23, Switzerland\\
$^3$Dipartimento di Fisica ``G.~Galilei'', Universit\`a di Padova and \\
INFN, Sezione di Padova, Via Marzolo~8, I-35131 Padua, Italy\\
\vspace{0.5cm}
$^*$Speaker. Talk given at the XLVIIIth Rencontres de Moriond session devoted to ELECTROWEAK INTERACTIONS AND UNIFIED THEORIES, La Thuile (Italy), 2-9 March 2013}

\maketitle\abstracts{\vspace*{0.5cm}
With the discovery of a scalar resonance at ATLAS and CMS, the understanding of the electroweak symmetry breaking origin seems a much closer goal. A strong dynamics at relatively low scales is still a good candidate. In this talk, the complete effective Lagrangian up to $d\leq5$ will be presented, both for the gauge and the flavour sectors. Interesting features in the flavour phenomenology will be discussed.}

\section{Framework}

With the new resonance at the Electroweak (EW) scale discovered at LHC~\cite{Aad:2012tfa,Chatrchyan:2012ufa}, we can now hope to have hints to understand the origin of the EW symmetry breaking (EWSB) mechanism. The data indicate that the new particle is looking more and more like the Standard Model (SM) scalar boson~\cite{Englert:1964et,Higgs:1964pj,Higgs:1964ia} with mass around 125 GeV, but other possibilities are still viable. In particular, the case of a strong dynamics at the TeV scale $\Lambda_s$ responsible for the EWSB is still attractive. In the original technicolor ansatz~\cite{Susskind:1978ms,Dimopoulos:1979es,Dimopoulos:1981xc}, only the three SM would-be-Goldstone bosons (GBs) are retained and are responsible of giving mass to the weak gauge bosons. Given the present discovery, a more attractive scenario is the so-called composite Higgs model, originally introduced in Ref.~\cite{Kaplan:1983fs,Kaplan:1983sm,Banks:1984gj,Georgi:1984ef,Georgi:1984af,Dugan:1984hq}. In this context, the theory is based on the spontaneous breaking of a large global symmetry, that gives rise to the appearance of several GBs: a realistic model then accounts for three GBs corresponding to the SM ones, and for at least one more GB that takes the role of the scalar boson. The latter is then a composite scalar degree of freedom that arises as massless and gets mass due to an explicit breaking of the symmetry. With respect to the technicolor context, here there are four relevant scales: $\Lambda_s$ typical of the strong resonances; $f\leq4\pi\Lambda_s$ that characterises the GBs energy scale; $v=246$ GeV defined through the $W$ mass, $M_W=g v/2$; and $\mean{h}$ that is the vacuum expectation value (VEV) of the scalar particle, providing the EWSB, and that is in general distinct from $v$. To measure the degree of non-linearity of these schemes, it is customary to introduce the parameter $\xi\equiv (v/f)^2$ that parametrises the impact of the strong dynamics at low-energy.

Without entering into details of a specific model, it is possible to describe the NP effects due to the TeV strong dynamics by making use of an effective Lagrangian approach, dealing with only the SM fields. The SM GBs can be described by a dimensionless unitary matrix:
\beq
\UH(x)=e^{i\sigma_a \pi^a(x)/v}\, , \qquad \qquad  \UH(x) \rightarrow L\, \UH(x) R^\dagger\,,
\nn
\eeq
with $L,R$ denoting respectively the $SU(2)_{L,R}$ global transformations of the scalar potential. The adimensionality of $\UH(x)$ is the technical key to understand why the dimension of the leading low-energy operators describing the dynamics of the scalar sector differs for a non-linear regime~\cite{Appelquist:1980vg,Longhitano:1980iz,Longhitano:1980tm,Feruglio:1992wf,Appelquist:1993ka} and a purely linear regime~\cite{Giudice:2007fh,Low:2009di}. In the former, non-renormalisable operators containing extra powers of a light $h$ are weighted by powers of $h/f$~\cite{Georgi:1984af}, and the GB contributions encoded in $\UH(x)$ do not exhibit any scale suppression. In the linear regime, instead, the light $h$ and the three SM GBs are encoded into the scalar doublet $H$, with mass dimension one: therefore any extra insertion of $H$ is suppressed by a power of the cutoff.

It is becoming customary to parametrise the Lagrangian describing a light dynamical scalar particle $h$ 
by means of the following ansatz~\cite{Contino:2010mh,Azatov:2012bz}
:
\begin{align}
\LL_h=&\phantom{+\,\,}\frac{1}{2} (\partial_\mu h) (\partial^\mu h) \,\left(1+c_H\,\xi\,\cF_H(h)\right) 
   \,-\, V(h) \,- \left(\frac{v}{2\sqrt{2}}\bar{Q}_L\UH(x)\,\cY\,\, Q_R \, \cF_Y(h)+\mbox{h.c.} \right)+\, 
\nn \\
& - \frac{v^2}{4}\Tr\left[\VL^\mu \VL_\mu\right] \,\cF_C(h) \,+\, 
    c_T\,\xi\,\frac{v^2}{4}\, \Tr\left[\TL\VL^\mu\right]\Tr\left[\TL\VL_\mu\right] \cF_T(h) \, + \, \ldots\,,
\label{L3}
\end{align}
where dots stand for higher order terms in the (linear) expansion in $h/f$, and \mbox{$\VL_\mu\equiv 
\left(\DL_\mu\UH\right)\UH^\dagger$} ($\TL\equiv\UH\sigma_3\UH^\dag$) is the vector (scalar) chiral field 
transforming in the adjoint of $SU(2)_L$. The covariant derivative reads 
\beq
\DL_\mu \UH(x) \equiv \derp_\mu \UH(x) +\dfrac{ig}{2}W_{\mu}^a(x)\sigma_a\UH(x) - 
                      \dfrac{ig'}{2} B_\mu(x) \UH(x)\sigma_3 \, , \nn
\eeq
with $W^a_\mu$ ($B_\mu$) denoting the $SU(2)_L$ ($U(1)_Y)$ gauge bosons and $g$ ($g'$) the corresponding 
gauge coupling. In the  equations above, $V(h)$ denotes the effective scalar potential describing the 
breaking of the EW symmetry. The first line in Eq.~(\ref{L3}) includes the SMS kinetic term, its scalar potential and the  Yukawa-like interactions for quarks, while the second line describes the $W$ and $Z$ masses and their interactions with $h$, as well as the usual custodial symmetry breaking term labeled by $c_T$. 

The functions $\cF_H(h)$, $\cF_C(h)$, $\cF_T(h)$ and $\cF_Y(h)$ above, as well as all $\cF(h)$ functions to be 
used below, encode the generic dependence on $(\mean{h}+h)$ and are model-dependent. Each $\cF(h)$ function can 
be expanded in powers of $\xi$, $\cF(h)= g_0(h,v) + \xi g_1(h,v) + \xi^2 g_2(h,v) + \ldots$, where $g(h,v)$ 
are model-dependent functions of $h$ and of $v$, once $\mean{h}$ is expressed in terms of $\xi$ and $v$. For not too small $\xi$ the whole series may need to be considered.

The above Lagrangian can be very useful to describe an extended class of ``Higgs'' models, ranging from 
the SM  scenario with a linear Higgs  sector (for $\mean{h}=v$, $a=b=c=1$ and neglecting the higher order terms 
in $h$), to the technicolor-like ansatz (for $f\sim v$ and omitting all terms in $h$) and intermediate 
situations with a light scalar $h$ (in general for $f\ne v$) as in composite/holographic Higgs models 
\cite{Kaplan:1983fs,Kaplan:1983sm,Banks:1984gj,Georgi:1984ef,Georgi:1984af,Dugan:1984hq,Agashe:2004rs,Contino:2006qr,Gripaios:2009pe} up to dilaton-like scalar frameworks. Note that in concrete models electroweak corrections imply $\xi< 0.2-0.4$~\cite{Contino:2010rs}, but we will leave the $\xi$ parameter free here and account for the constraints on custodial symmetry through limits on the $d=2$ and higher-dimensional chiral operator coefficients.

In what follows, the complete basis of independent operators up to dimension 5 will be reported, both in the gauge and in the flavour sectors, providing the complete list of interactions of a light $h$~\cite{Alonso:2012jc,Alonso:2012px,Alonso:2012pz}. This analysis enlarges and completes the operator basis previously considered in Refs.~\cite{Appelquist:1980vg,Longhitano:1980iz,Longhitano:1980tm,Feruglio:1992wf,Appelquist:1993ka,Contino:2010mh,Azatov:2012bz} and represents a fundamental tool in order to characterise the emerging phenomenology at LHC and investigating on the EWSB origin.

\section{The effective Lagrangian in the gauge sector}

All CP-even gauge operators appropriate to the non-linear regime will be included in this section, up to mass 
dimension $5$. In the absence of a light $h$, no pure gauge or gauge-$h$ $d=5$ operator exists, 
and it is thus a good guideline to start from the basis of $d=4$ pure gauge chiral operators and complete 
it up to $d=5$ with suitable insertions of $h$. The connection 
to the linear regime will be made manifest exploiting the operator dependence on $\xi$. The Lagrangian can be 
decomposed as 
\beq
\begin{split}
\LL^{d\le5}_{gauge-h}= &
\LL_h-\frac{g_s^2}{4}\,G_{\mu\nu}^aG^{\mu\nu}_a\,\cF_G(h)
-\frac{g^2}{4}\,W_{\mu\nu}^aW^{\mu\nu}_a\,\cF_W(h)
-\frac{g'^2}{4}\,B_{\mu\nu}B^{\mu\nu}\,\cF_B(h)+ \\
&+\xi\, \sum_{i=1}^{5} \,c_i\,\cP_i(h)\,
+\, \xi^2 \,\sum_{i=6}^{20}\, c_i\,\cP_i(h)
+\, \xi^3 \,\sum_{i=21}^{23}\, c_i\,\cP_i(h)
+\, \xi^4 \, c_{24}\,\cP_{24}(h) \,.
\end{split}
\label{L4}
\eeq
The first line of Eq.~(\ref{L4}) contains the kinetic terms for the gauge bosons, with $W_{\mu\nu}$, 
$B_{\mu\nu}$ and $G_{\mu\nu}$ denoting the $SU(2)_L$, $U(1)_Y$ and $SU(3)_C$ field strengths, respectively. 
The second line of Eq.~(\ref{L4}) contains the following $24$ CP-even operators, ordered by their $\xi$ dependence~\cite{Alonso:2012px}:
\bea
&
\begin{aligned}
\blue{\cP_{1}(h)}\,\,  &= g\,g' \,B_{\mu\nu} \Tr\left(\TL\,W^{\mu\nu}\right)\,\cF_{1}(h)
&\red{\cP_{4}(h)}\,\,  &= i\,g' B_{\mu\nu}\Tr(\TL\VL^\mu)\,\derp^\nu \cF_{4}(h)\\
\blue{\cP_{2}(h)}\,\,  &= i\,g' \,B_{\mu\nu} \Tr\left(\TL\left[\VL^\mu,\VL^\nu\right]\right)\,\cF_{2}(h) 
&\red{\cP_{5}(h)}\,\,  &= i\,g \,\Tr(W_{\mu\nu}\VL^\mu)\,\derp^\nu \cF_{5}(h) \\
\blue{\cP_{3}(h)}\,\,  &= i\,g\,\Tr\left(W_{\mu\nu}\left[\VL^\mu,\VL^\nu\right]\right)\,\cF_{3}(h)
\end{aligned}
\label{GaugeOperators1d5}\\
\nonumber\\
&
\begin{aligned}
\blue{\cP_{6}(h)}\,\, &=\left(\Tr\left(\VL_\mu\,\VL^\mu\right)\right)^2\,\cF_{6}(h) 
&\red{\cP_{14}(h)} &= i\,g \,\Tr(\TL W_{\mu\nu})\Tr(\TL\VL^\mu)\,\derp^\nu \cF_{14}(h)\\
\blue{\cP_{7}(h)}\,\, &=\left(\Tr\left(\VL_\mu\,\VL_\nu\right)\right)^2\,\cF_{7}(h)
&\red{\cP_{15}(h)} &=\Tr(\TL\,[\VL_\mu,\VL_\nu])\Tr(\TL\VL^\mu)\,\derp^\nu \cF_{15}(h)\\
\blue{\cP_{8}(h)}\,\, &=g^2\,\left(\Tr\left(\TL\,W^{\mu\nu}\right)\right)^2\,\cF_{8}(h) 
&\red{\cP_{16}(h)} &=\Tr(\VL_\nu \,\cD_\mu\VL^\mu)\,\derp^\nu \cF_{16}(h)\\
\blue{\cP_{9}(h)}\,\, &=i\,g\,\Tr\left(\TL\,W_{\mu\nu}\right)\Tr\left(\TL\left[\VL^\mu,\VL^\nu\right]\right)\,
                       \cF_{9}(h)
&\red{\cP_{17}(h)} &=\Tr(\TL\,\cD_\mu\VL^\mu)\Tr(\TL\VL_\nu)\,\derp^\nu \cF_{17}(h) \\
\blue{\cP_{10}(h)} &=g\,\epsilon^{\mu\nu\rho\lambda}\Tr\left(\TL\VL_\mu\right)\Tr\left(\VL_\nu\,W_{\rho\lambda}
                       \right)\,\cF_{10}(h)
&\red{\cP_{18}(h)} &=\Tr\left(\VL_\mu\,\VL^\mu\right)\,\derp_\nu\derp^\nu\cF_{18}(h)\\
\green{\cP_{11}(h)} &= \Tr\left((\cD_\mu\VL^\mu)^2 \right)\,\cF_{11}(h)
&\red{\cP_{19}(h)} &=\Tr\left(\VL_\mu\,\VL_\nu\right)\,\derp^\mu\cF_{19}(h)\derp^\nu\cF'_{19}(h)\\
\green{\cP_{12}(h)} &= \Tr(\TL\,\cD_\mu\VL^\mu)\,\Tr(\TL\,\cD_\nu\VL^\nu)\,\cF_{12}(h)
&\red{\cP_{20}(h)} &=\Tr\left(\TL\VL_\mu\right)\Tr\left(\TL\VL_\nu\right)\,\derp^\mu\cF_{20}(h)\derp^\nu\cF'_{20}(h)\\
\green{\cP_{13}(h)} & = \Tr([\TL \,,\VL_\nu]\,\cD_\mu \VL^\mu) \, \Tr(\TL\VL^\nu) \cF_{13}(h) \\
\end{aligned} 
\label{GaugeOperators6d20} \\ 
\nn \\
&
\begin{aligned}
\blue{\cP_{21}(h)} &= \Tr\left(\VL_\mu\VL^\mu\right)\left(\Tr\left(\TL\VL_\nu\right)\right)^2\cF_{21}(h)
&\red{\cP_{23}(h)} &=\left(\Tr\left(\TL\,\VL_\mu\right) \right)^2 \derp_\nu\derp^\nu\cF_{23}(h)\\
\blue{\cP_{22}(h)} &=\Tr\left(\VL_\mu\VL_\nu\right)\Tr\left(\TL\VL^\mu\right)\Tr\left(\TL\VL^\nu\right)\cF_{22}(h)\\
\end{aligned}
\label{GaugeOperators21d23}\\
\nonumber\\
&
\begin{aligned}
\blue{\cP_{24}(h)} &=\left(\Tr\left(\TL\VL_\mu\right)\Tr\left(\TL\VL_\nu\right)\right)^2\,\cF_{24}(h)\,.
\end{aligned}
\label{GaugeOperators24}
\eea
The 24 constant parameters $c_i$ are model-dependent coefficients. The powers of $\xi$, factorized out in 
the second line of Eq.~(\ref{L4}), do not reflect an expansion in $\xi$, but a reparametrisation that facilitates the tracking to the lowest dimension at which a ``sibling'' operator appears in the linear expansion. By sibling we mean an operator written in terms of the scalar doublet $H$, that includes the pure gauge part of the couplings $\cP_{1-24}(h)$. It may happen that an operator listed in Eqs.~(\ref{GaugeOperators1d5})-(\ref{GaugeOperators24}) corresponds to a specific 
combination of siblings with different dimensions. This is the case, for instance, of $\cP_{13}(h)$, 
whose siblings are of dimension $8$ and $10$.

For $\xi\ll 1$ the weight of the operators which are accompanied by powers of $\xi$ is scale suppressed 
compared to that of SM renormalisable couplings. In this limit the Lagrangian above would encode a consistent 
linear expansion up to $d=6$ operators, if only the terms of zero and first order in $\xi$ are kept: indeed, 
operators  $\cP_{6}(h)$ to $\cP_{24}(h)$ would correspond to $d=8$ or higher-dimension siblings in the linear 
expansion. In contrast, in the non-linear regime, that is for $\xi\approx 1$, no such suppression appears 
and {\it all} operators in Eqs.~(\ref{GaugeOperators1d5})-(\ref{GaugeOperators24}) include $d \le 5$ 
couplings and should be considered on equal footing. The leading terms of the linear and non-linear 
expansions do not match.

The different operators defined in Eqs.~(\ref{GaugeOperators1d5})-(\ref{GaugeOperators24}) correspond 
to three major categories: pure gauge and gauge-$h$ operators (in blue) which result from a direct extension of the original Appelquist-Longhitano chiral Higgsless  basis; operators containing the contraction 
$\cD_\mu \VL^\mu$ and no derivatives of $\cF(h)$ (in green); operators with one or two derivatives of 
$\cF(h)$ (in red).

\section{The effective Lagrangian in the flavour sector}

The core of the flavour problem in NP theories consists in explaining the high level of suppression that must 
be encoded in most of the theories beyond the SM in order to pass flavour changing neutral current (FCNC) 
observability tests. Minimal Flavour Violation (MFV)~\cite{Chivukula:1987py,Hall:1990ac,D'Ambrosio:2002ex,Cirigliano:2005ck,Davidson:2006bd,Kagan:2009bn,Gavela:2009cd,Feldmann:2009dc,Alonso:2011yg,Alonso:2011jd,Alonso:2012fy} emerged in the last years as one of the most promising working frameworks to this end\footnote{Notice that the MFV should be consider just a counting rule and not a model of flavour as there is no explanation of the origin of fermion masses and mixing, or equivalently there is no explanation of the background values of the Yukawa spurions~\cite{Alonso:2012jc,Alonso:2012fy}. With this respect, more successful contexts have been constructed dealing with symmetries smaller than the MFV one~\cite{Altarelli:2005yp,Altarelli:2005yx,Feruglio:2007uu,Bazzocchi:2009pv,Bazzocchi:2009da,Altarelli:2009gn,Altarelli:2009kr,Toorop:2010yh,Meloni:2011fx,Varzielas:2010mp,Altarelli:2012bn,Altarelli:2012ss,Bazzocchi:2012st,Altarelli:2012ia}. It turns out that also for these models, the scale of NP can be lowered down to the TeV scale~\cite{Feruglio:2008ht,Feruglio:2009iu,Feruglio:2009hu,Merlo:2011hw}.}: the MFV ansatz dictates that flavour in the SM and beyond is described at low-energies uniquely in terms of the known fermion mass hierarchies and mixings. An outcome is that the energy scale of the NP may be as low as few TeV in several distinct contexts \cite{Lalak:2010bk,Fitzpatrick:2007sa,Grinstein:2010ve,Buras:2011wi,Lopez-Honorez:2013wla}, while in general it should be larger than hundreds of TeV~\cite{Isidori:2010kg}. 

In Ref.~\cite{D'Ambrosio:2002ex}, the complete basis of gauge-invariant 
6-dimensional FCNC operators has been constructed for the case of a linearly realized SM Higgs sector, 
 in terms of the SM fields and the $Y_U$ and $Y_D$ spurions. Operators of dimension $d>6$ are usually 
neglected due to the additional suppression in terms of the cut-off scale. In what follow we present the corresponding analysis in the case of a strong dynamics at the TeV scale and discuss some phenomenological features.

\subsection{$d=4$ chiral operators}
In the non-linear regime a chiral expansion is pertinent, and this results in a different set of operators at leading order than in the case of the linear regime. A total of four independent $d=4$ chiral operators containing LH fermion fields can be constructed~\cite{Alonso:2012jc,Appelquist:1984rr,Cvetic:1988ey,Espriu:2000fq}, namely:
\bea
&\LL_{\chi=4}^f=\xi \sum_{i=1,2, 3} \hat{a}_i\,\cO_i(h)+\xi^2  \hat{a}_4\,\cO_4(h)
\label{d4OperatorsH}\\
&\begin{aligned}
&\mathcal{O}_{1}(h)=\frac{i}{2}\,\bar{Q}_L\,\lambda_{F}\,\gamma^{\mu}\,\left\{\TL,\VL_{\mu} \right\}\, Q_L\,\cF_1(h)\,, 
&\qquad & \mathcal{O}_{2}(h)=i\,\bar{Q}_L\,\lambda_{F}\,\gamma^{\mu}\,\VL_{\mu}\,Q_L\,\cF_2(h)\,, \\
&\mathcal{O}_{3}(h)=i\,\bar{Q}_L\,\lambda_{F}\,\gamma^{\mu}\,\TL\,\VL_{\mu}\,\TL\, Q_L\,\cF_3(h)\,, & \qquad &
\mathcal{O}_{4}(h)=\frac{1}{2}\,\bar{Q}_L\,\lambda_{F}\,\gamma^{\mu}\,\left[\TL,\VL_{\mu}\right]\,Q_L\,\cF_4(h)\,,
\end{aligned}
\label{d4Operators}
\eea
where the parameter $\lambda_F$ remembers the MFV ansatz,
\beq
\lambda_F\equiv Y_U\,Y_U^\dag+Y_D\,Y_D^\dag=V^\dag\cy_U^2 V+\cy_D^2.
\eeq
Out of these $\cO_1(h)-\cO_3(h)$ are CP-even while $\cO_4(h)$ is intrinsically CP-odd \cite{Alonso:2012jc}. The powers of $\xi$ in Eq.~(\ref{d4OperatorsH}) facilitate the identification of the lowest dimension at which a sibling operator appears in the linear regime. The lowest-dimension siblings of $\cO_1(h)-\cO_3(h)$ arise at $d=6$, while that of $\cO_4$ appears at $d=8$~\cite{Alonso:2012jc}. 

Operators $\cO_1(h)-\cO_3(h)$ induce tree-level contributions to $\Delta F=1$ processes mediated by the $Z$ boson and are severely constrained. Due to the MFV structure of the coefficients, sizable flavour-changing effects may only be expected in the  down quark sectors, with data on $K$ and $B$ transitions providing the strongest constraints on   $a_Z^d$,   
\beq
-0.044<a_Z^d<0.009\qquad\qquad \text{at $95\%$ of C.L.}
\eeq
from $K^+\to\pi^+\bar\nu\nu$, $B\to X_s\ell^+\ell^-$ and $B\to\mu^+\mu^-$ data.

Furthermore, operators $\cO_2(h)-\cO_4(h)$ induce corrections to the fermion-$W$ couplings, and thus to the  the CKM matrix. This in turn induces modifications~\cite{Alonso:2012jc} on the strength of meson oscillations (at loop level), on $B^+\to\tau^+\nu$ decay and on the $B$ semileptonic CP-asymmetry, among others; more specifically the following process have been taken into account in Ref.~\cite{Alonso:2012jc}:
\begin{itemize}
\item[-] The CP-violating parameter $\epsilon_K$ of the $K^0-\bar K^0$ system and the mixing-induced CP asymmetries $S_{\psi K_S}$ and $S_{\psi\phi}$ in the decays $B^0_d\to\psi K_S$ and $B^0_s\to\psi\phi$. Possible large deviations from the values predicted by the SM  are only allowed in the $K$ system.

\item[-] The ratio among the meson mass differences in the $B_d$ and $B_s$ systems, $R_{\Delta M_B}\equiv \Delta M_{B_d}/\Delta M_{B_s}$. Deviations from the SM prediction for this observable are negligible.

\item[-] The ratio among the $B^+\to\tau^+\nu$ branching ratio and the $B_d$ mass difference, $R_{BR/\Delta M}\equiv BR(B^+\to\tau^+\nu)/\Delta M_{B_d}$. This observable is clean from theoretical hadronic uncertainties.

\item[-] The $\bar B\to X_s\gamma$ branching ratio that benefits of good experimental and theoretical precision.
\end{itemize}
Since only small deviations from the SM prediction for $S_{\psi K_S}$ are allowed,  only values close to the exclusive determination for $|V_{ub}|$ are favoured. Moreover, it is possible to constrain the $|V_{ub}|-\gamma$ parameter space, with $\gamma$ being one of the angles of the unitary triangle, requiring that both $S_{\psi K_S}$ and $R_{\Delta M_B}$ observables are inside the $3\sigma$ experimental determination. 

\begin{figure}[h!]
 \centering
 \subfigure[Correlation plot between $\vep_K$ and $R_{BR/\Delta M}$. $a_W,a_{CP}\in{[}-1,1{]}$, 
            $a_Z^d\in {[}-0.1,0.1{]}$]{
\includegraphics[width=7.6cm]{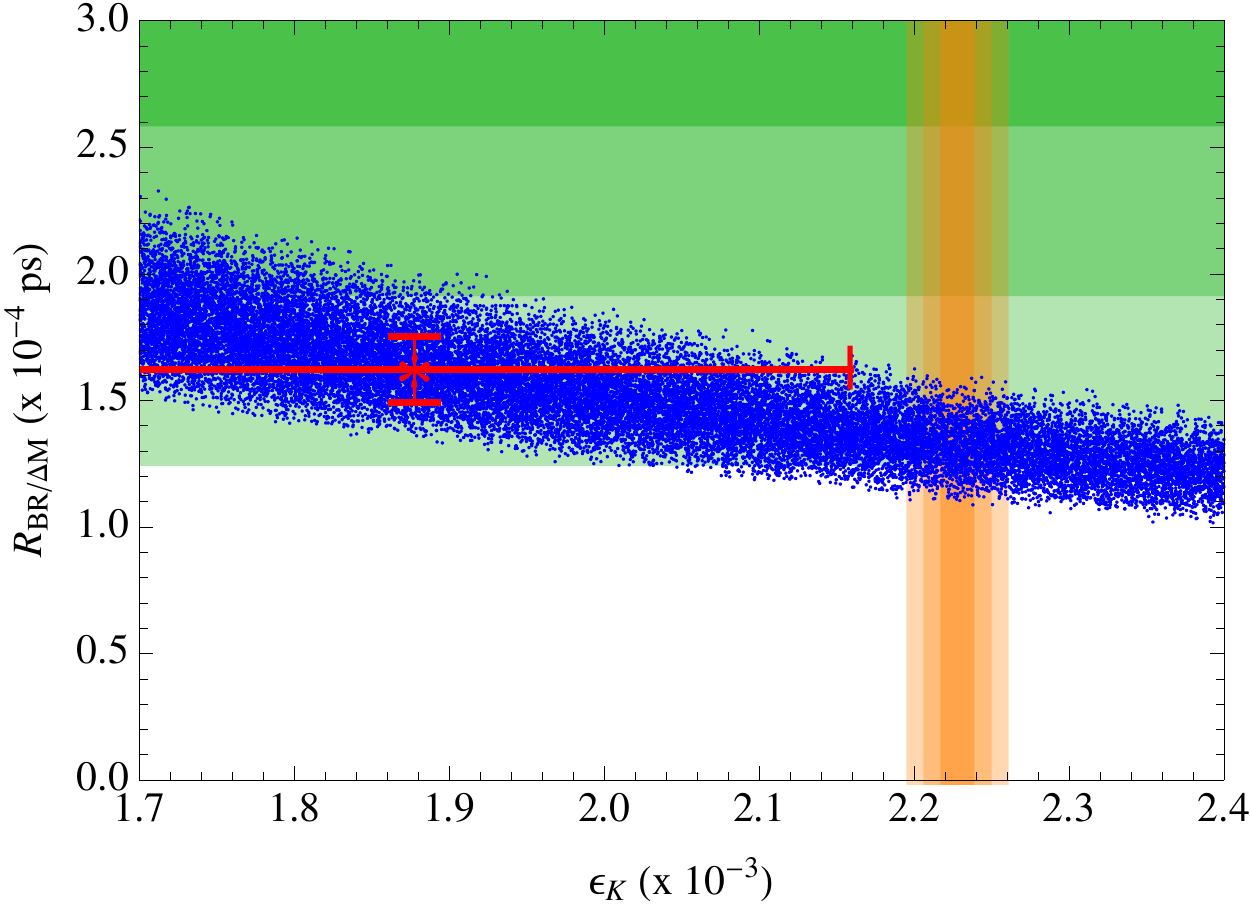}}  
\subfigure[$a_W-a_{CP}$ parameter space for the observables on the left panel inside their $3\sigma$ error ranges and 
            $a_Z^d\in{[}-0.044,0.009{]}$.]{
\includegraphics[width=7.8cm]{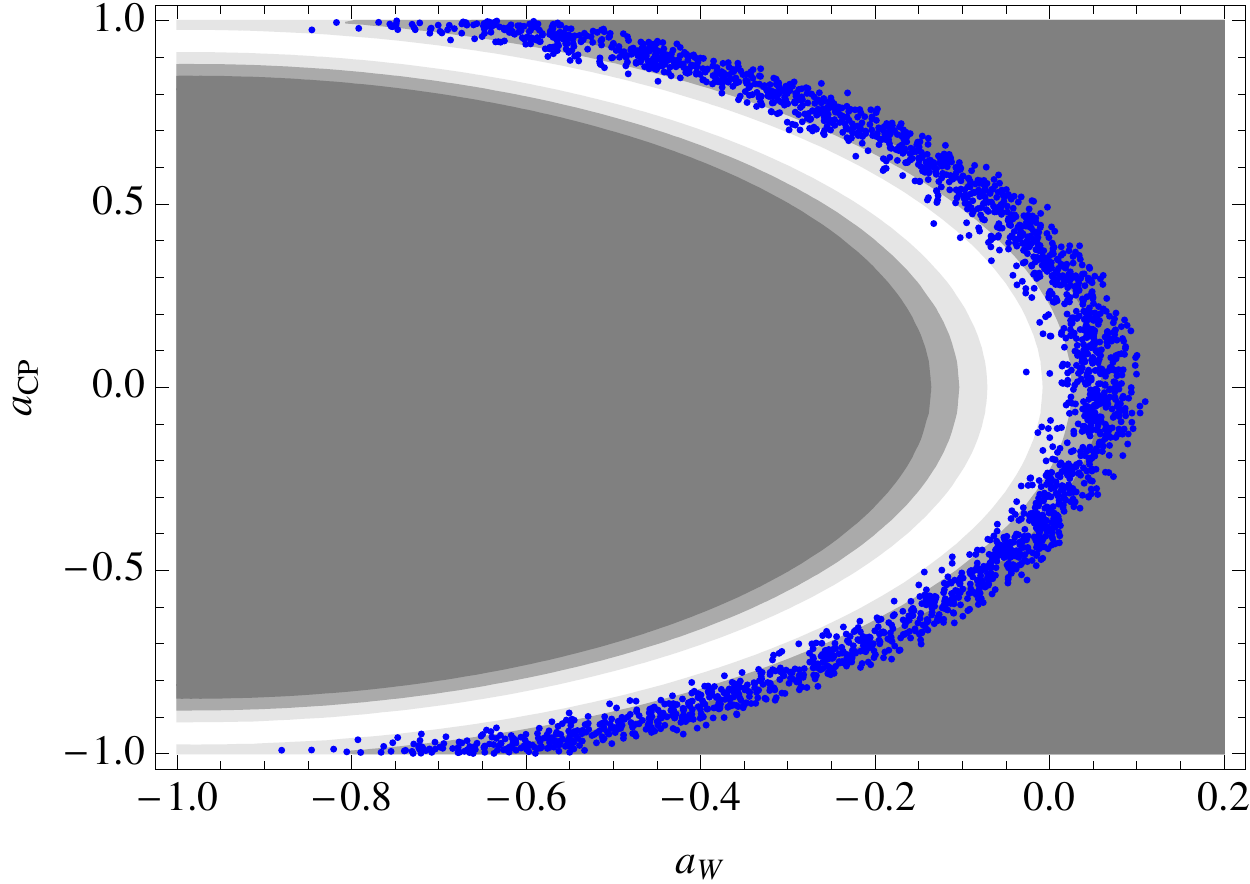}}
\caption{Results for the reference point $(|V_{ub}|,\,\gamma)=(3.5\times 10^{-3},\,66^\circ)$. See the text for details.}
\label{fig:EpsilonKRatio}
\end{figure}

Once  this reduced parameter space is identified, it is illustrative to choose one of its points as reference point, in order to present the features of this MFV scenario; for instance for the values $(|V_{ub}|,\gamma)=(3.5\times 10^{-3}, 66^\circ)$, $S_{\psi K_S}$, $R_{\Delta M_B}$ and $|V_{ub}|$ are all inside their own $1\sigma$ values, and the predicted SM values for $\epsilon_K$ and $R_{BR/\Delta M}$ are
\beq
\epsilon_K=1.88\times10^{-3}\,,\qquad\qquad
R_{BR/\Delta M}=1.62\times 10^{-4}\,.
\eeq
The errors on these quantities are  $\sim15\%$ and $\sim8\%$, estimated considering the uncertainties on the input parameters and the analysis performed in Ref.~\cite{Brod:2011ty}.
Fig.~\ref{fig:EpsilonKRatio} shows the correlation between $\epsilon_K$ and $R_{BR/\Delta M}$ (left panel) and the $a_{CP}-a_W$ parameter space (right panel), requiring that $\epsilon_K$ and $R_{BR/\Delta M}$ lie inside their own $3\sigma$ experimental determination. In the latter, the gray areas correspond to the bounds from the $BR(\bar B\to X_s\gamma)$. Finally, for those points in the $a_{CP}-a_W$ parameter space that pass all the previous constraints, the predictions for $S_{\psi\phi}$ and the $B$ semileptonic CP-asymmetry turned out to be close to the SM determination, in agreement with the recent LHCb measurements~\cite{LHCb:2011aa}.

Fig.~\ref{fig:EpsilonKRatio} on the right  shows that $a_{CP}$, the overall coefficient of the genuinely CP-odd coupling $\cO_4(h)$ is still loosely constrained by low-energy data. This has an interesting phenomenological consequence on Higgs physics prospects, since it translates into correlated exotic Higgs-fermion couplings, which for instance at leading order in $h$ read: 
\beq
\delta \LL_{\chi=4}^h \supset a_{CP}\left(1 + \beta_{CP}\frac{h}{v}\right)\,\cO_4\,\,.
\eeq
These are encouraging results in the sense of allowing short-term observability. In  a conservative perspective, the operator coefficients of the $d=4$ non-linear expansion should be expected to be $\cO(1)$. Would this be the case, the possibility of NP detection would be delayed until both low-energy flavour experiments and LHC precision on  $h$-fermion couplings nears the $\cO(10^{-2})$ level, which for LHC means  to reach at least its $3000\,fb^{-1}$ running regime. Notwithstanding this, a steady improvement of the above bounds should be sought.

\subsection{$d=5$ chiral operators}
Gauge invariant $d=5$ operators relevant for flavour can be classified in three main groups: 
\begin{description}
\item[i) dipole-type operators:]
\beq
\begin{aligned}
&\mathcal{X}_{1}=  g'\,\bar{Q}_L\,\tmnp\,\UH\,Q_R \,B_{\mu\nu}\,,\qquad\qquad
&&\mathcal{X}_{2}= g' \,\bar{Q}_L\,\tmnp\,\TL\,\UH\,Q_R\,B_{\mu\nu}\,,\\
&\mathcal{X}_{3}=  g \,\bar{Q}_L\,\tmnp\,\sigma_i\UH\,Q_R\,W^i_{\mu\nu}\,,\qquad\qquad
&&\mathcal{X}_{4}= g \,\bar{Q}_L\,\tmnp\,\sigma_i\TL\,\UH\,Q_R\,W^i_{\mu\nu}\,,\\
&\mathcal{X}_{5}=  g_s \,\bar{Q}_L\,\tmnp\,\UH\,Q_R\,G_{\mu\nu}\,,\qquad\qquad
&&\mathcal{X}_{6}= g_s \, \bar{Q}_L\,\tmnp\,\TL\,\UH\,Q_R\,G_{\mu\nu}\,, \\
&\mathcal{X}_{7}=  g \,\bar{Q}_L\,\tmnp\,\TL\,\sigma_i\,\UH\,Q_R\,W^i_{\mu\nu}\,,\qquad\qquad
&&\mathcal{X}_{8}= g \,\bar{Q}_L\,\tmnp\,\TL\,\sigma_i\TL\,\UH\,Q_R\,W^i_{\mu\nu}\,;
\end{aligned}
\label{OpGroup3}
\eeq
\item[ii) operators containing the rank-2 antisymmetric tensor $\tmnp$:]
\beq
\begin{aligned}
&\mathcal{X}_{9\phantom{0}}=\bar{Q}_L\,\tmnp\,[\VL_\mu,\VL_\nu]\,\UH\,Q_R\,,\qquad
&&\mathcal{X}_{10}=\bar{Q}_L\,\tmnp\,[\VL_\mu,\VL_\nu]\,\TL\,\UH\,Q_R\,,\\
&\mathcal{X}_{11}=\bar{Q}_L\,\tmnp\,[\VL_\mu \,\TL,\VL_\nu \,\TL]\,\UH\,Q_R\,,\qquad
&&\mathcal{X}_{12}=\bar{Q}_L\,\tmnp\,[\VL_\mu \,\TL,\VL_\nu \,\TL]\,\TL\,\UH\,Q_R\,;
\end{aligned}
\label{OpGroup2}
\eeq
\item[iii) other operators containing the chiral vector fields $\VL_\mu$:]
\beq
\begin{aligned}
&\mathcal{X}_{13}=\bar{Q}_L\,\VL_\mu\, \VL^\mu\,\UH \,Q_R\,,\qquad\qquad
&&\mathcal{X}_{14}=\bar{Q}_L\, \VL_\mu\, \VL^\mu\, \TL\,\UH \,Q_R\,,\\
&\mathcal{X}_{15}=\bar{Q}_L\, \VL_\mu\, \TL\, \VL^\mu\,\UH \,Q_R\,,\qquad\qquad 
&&\mathcal{X}_{16}=\bar{Q}_L\, \VL_\mu\,\TL \,\VL^\mu\,\TL\,\UH \,Q_R\,,\\
&\mathcal{X}_{17}=\bar{Q}_L\, \TL \,\VL_\mu \,\TL \,\VL^\mu\,\UH \,Q_R\,,\qquad\qquad 
&&\mathcal{X}_{18}=\bar{Q}_L\, \TL\, \VL_\mu \,\TL \,\VL^\mu \,\TL\,\UH \,Q_R\,. 
\end{aligned}
\label{OpGroup1}
\eeq
\end{description}

The chiral Lagrangian containing the 18 fermionic flavour-changing $d=5$ operators can thus be written as
\beq
\LL_{\chi=5}^f= \sqrt{\xi}\,\sum_{i=1}^{8}\,b_i\,\dfrac{\cX_i}{\Lambda_s}+\xi\sqrt{\xi}\sum_{i=9}^{18}\,b_i\,\dfrac{\cX_i}{\Lambda_s}\,. 
\label{OrigLag5-bis}
\eeq
where $\Lambda_s$ is the scale of the strong dynamics and $b_i$ are arbitrary $\cO(1)$ coefficients. It is worth to underline that for the analysis of $d=5$ operators in the non-linear regime, the relevant scale is $\Lambda_s$ and not $f$ as for the analysis in the previous section. Indeed, $f$ is associated to light Higgs insertions, while $\Lambda_s$ refers to the characteristic scale of the strong resonances that, once integrated out, give rise to the operators listed in Eqs.~(\ref{OpGroup3})-(\ref{OpGroup1}). In the limit of small $\xi$, $\cX_{1-6}$ correspond to $d=6$ operators in the linear expansion, while $\cX_{7}$ and $\cX_{8}$ result from combinations of $d=6$ and $d=8$ siblings. Moreover, $\cX_{9-18}$ have linear siblings of $d=8$, but $\cX_{17}$ and $\cX_{18}$ that are combinations of $d=8$ and $d=10$ operators in the linear regime. 

\begin{figure}[h!]
 \centering
\includegraphics[width=7.6cm]{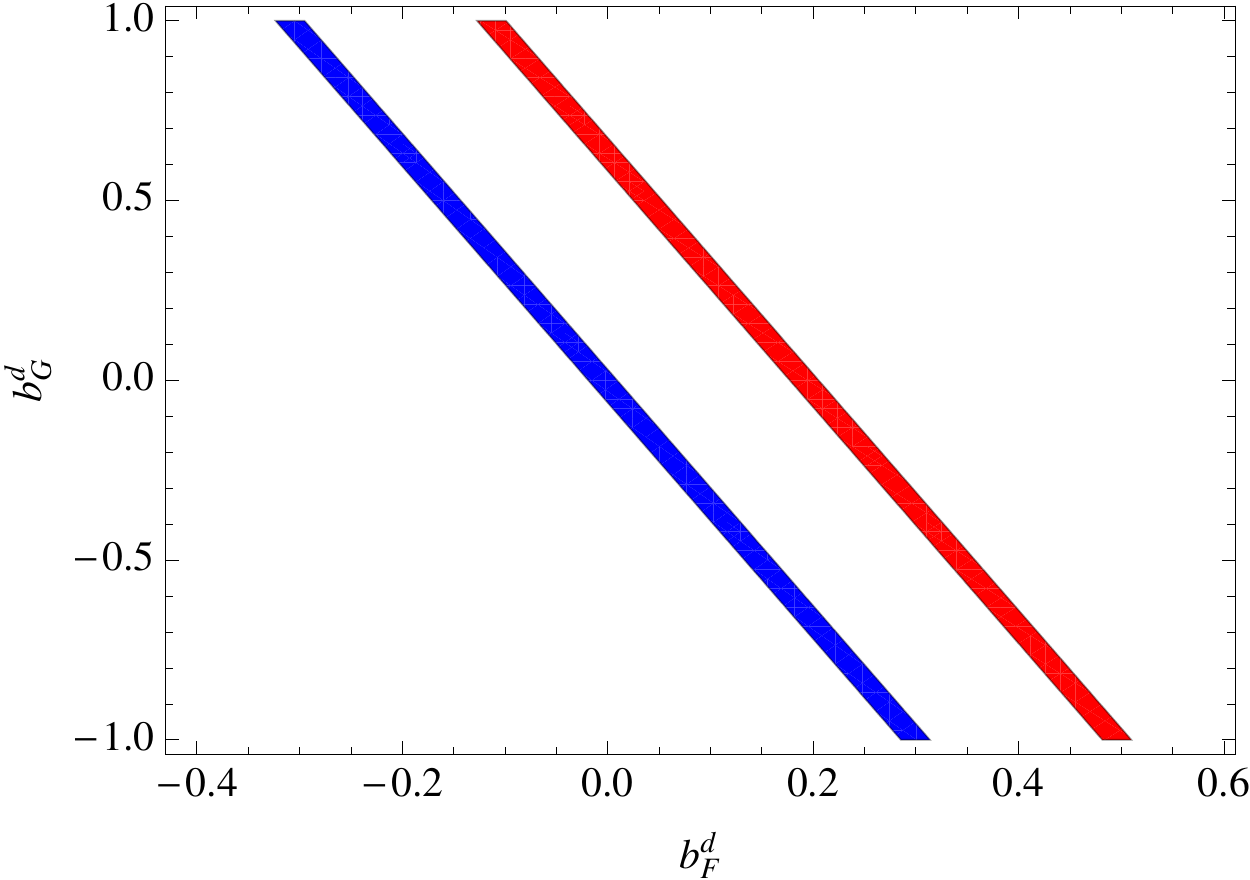}
\caption{\it The $BR(\bar B\to X_s\gamma)$ $3\sigma$-allowed bands in the $b^d_F-b^d_G$ parameter space.}
\label{fig:BSG5f4pv}
\end{figure}

The phenomenological impact of these contributions can be best identified through the low-energy Lagrangian 
written in the unitary gauge: in the following we will concentrate only on the dipole operators 
\beq
e\dfrac{d_F^d}{\Lambda_s}\bar D_L \,\tmnp D_R F_{\mu\nu}\,,\qquad\qquad
+g_s\dfrac{d_G^d}{\Lambda_s}\bar D_L \,\tmnp D_R G_{\mu\nu}\,,
\eeq
that have an interesting impact on $BR(\bar B\to X_s\gamma)$ (a complete discussion can be found in Ref.~\cite{Alonso:2012pz}). Considering the experimental determination and the theoretical prediction for $BR(\bar B\to X_s\gamma)$, it is possible to constrain the $b^d_F-b^d_G$ parameter space and the result are shown in Fig.~\ref{fig:BSG5f4pv}, where the two narrow bands depict the two allowed regions. 

Analogously to the case of $\cO_1(h)\dots \cO_4(h)$ operators discussed in the previous subsection, a correlation would hold between a low-energy signal from these $d=5$ couplings and the detection of exotic fermionic couplings at LHC, upon considering their extension  to include $h$-dependent insertions.

\section*{Acknowledgments}
We acknowledge partial support by European Union FP7 ITN INVISIBLES (Marie Curie Actions, PITN-GA-2011-289442), 
CiCYT through the project FPA2009-09017, CAM through the project HEPHACOS P-ESP-00346, European Union FP7 
ITN UNILHC (Marie Curie Actions, PITN-GA-2009-237920), MICINN through the grant BES-2010-037869 and the Juan 
de la Cierva programme (JCI-2011-09244), Italian Ministero 
dell'Uni\-ver\-si\-t\`a e della Ricerca Scientifica through the COFIN program (PRIN 2008) and the contracts 
MRTN-CT-2006-035505 and  PITN-GA-2009-237920 (UNILHC). Finally we thank the organizers of the Moriond EW conference for the kind invitation and for their efforts in organizing this enjoyable meeting.

\end{document}